\documentclass[%
 reprint,
superscriptaddress,
 amsmath,amssymb,
 aps,
]{revtex4-1}

\usepackage{graphicx}
\usepackage{dcolumn}
\usepackage{bm}
\usepackage{hyperref}
\usepackage{color}
\usepackage{ulem}

\def \e {\varepsilon}

\newcommand {\vishnu}   {{\tt iEBE-VISHNU}}
\newcommand{\trento}{TRENTo}

\begin{document}

\preprint{APS/123-QED}

\title{Collective flow and the fluid behavior in p/d/$^3$He+Au collisions at $\sqrt{s_{NN}} = 200$~GeV}%

\author{Zeming Wu}
\affiliation{School of Physics, Peking University, Beijing 100871, China}
\affiliation{Collaborative Innovation Center of Quantum Matter, Beijing 100871, China}

\author{Baochi Fu}%
\email{fubaochi@pku.edu.cn}
\affiliation{Center for High Energy Physics, Peking University, Beijing 100871, China}
\affiliation{School of Physics, Peking University, Beijing 100871, China}
\affiliation{Collaborative Innovation Center of Quantum Matter, Beijing 100871, China}

\author{Shujun Zhao}
\affiliation{School of Physics, Peking University, Beijing 100871, China}
\affiliation{Collaborative Innovation Center of Quantum Matter, Beijing 100871, China}

\author{Runsheng Liu}%
\affiliation{School of Physics, Peking University, Beijing 100871, China}
\affiliation{Collaborative Innovation Center of Quantum Matter, Beijing 100871, China}

\author{Huichao Song}
\email{huichaosong@pku.edu.cn}
\affiliation{School of Physics, Peking University, Beijing 100871, China}
\affiliation{Collaborative Innovation Center of Quantum Matter, Beijing 100871, China}
\affiliation{Center for High Energy Physics, Peking University, Beijing 100871, China}
\date{\today}

\begin{abstract}
By varying the intrinsic initial geometry, the p/d/$^3$He+Au collisions at the Relativistic Heavy Ion Collider (RHIC) provide a unique opportunity to understand the collective behavior and probe the possible sub-nucleon fluctuations in small systems.
In this paper, we employ the hybrid model \vishnu~ with \trento~ initial conditions to study the collective flow and the fluid behavior in p/d/$^3$He+Au collisions. With fine-tuned  parameters, \vishnu~ can describe the $v_2(p_T)$ and $v_3(p_T)$ data from the PHENIX and STAR collaborations. However, for some certain parameter sets with initial sub-nucleon fluctuation, the hydrodynamic simulations have already beyond their limits with the average Knudsen number $\langle K_n \rangle$ obviously larger than one. Our calculations demonstrate that, for a meaningful evaluation of the fluid behavior in the small systems, model simulations should also pay attention to the validity range of hydrodynamics.\end{abstract}

\maketitle

\section{\label{sec:intro}Introduction}
Over the last two decades, the properties of the extremely hot and dense QCD matter, the quark-gluon plasma (QGP), has been studied intensively by the relativistic heavy-ion programmes at the Relativistic Heavy Ion Collider (RHIC) and the Large Hadron Collider (LHC). It has been found that the QGP created in large collision systems, such as
Au+Au and Pb+Pb collisions, behaves as an almost ``perfect liquid"~\cite{Kolb:2003dz,PHENIX:2004vcz,BRAHMS:2004adc,PHOBOS:2004zne,Gyulassy:2004zy,STAR:2005gfr,Muller:2006ee,Jacak:2012dx,Muller:2012zq,Gale:2013da,Heinz:2013th,Shuryak:2014zxa,Song:2013gia,Song:2017wtw,Busza:2018rrf}. Its strong collective expansion and associated flow observables have been successfully described by hydrodynamic calculations with small specific shear viscosity, which is close to the {lowest KSS bound} \cite{Song:2010mg,Bernhard:2019bmu,JETSCAPE:2020shq}. The small collision systems, such as the p+Pb and p+p collisions at the LHC, were originally intended to provide the reference data for the large collision systems. However, similar collective behaviour has been observed in the high multiplicity events, including the ``double ridge'' structure in two particle correlations\cite{CMS:2012qk,CMS:2013pdl,CMS:2013jlh,CMS:2015yux,CMS:2015xmx,ATLAS:2013jmi,ATLAS:2014qaj,ATLAS:2012cix,ALICE:2012eyl,ALICE:2013snk,ALICE:2014dwt}, multi-particle culumants \cite{ATLAS:2013jmi,CMS:2015yux,ATLAS:2016yzd,CMS:2016fnw,ALICE:2019zfl,ALICE:2014dwt}, and the mass ordering of the anisotropic flow  of identified particles \cite{CMS:2016fnw,ALICE:2013snk,CMS:2014und}, see \cite{Song:2017wtw,Loizides:2016tew,Schlichting:2016sqo,Nagle:2018nvi} for recent review.

To understand these flow-like observables, a key question is whether the QGP droplets were formed in the small collision systems. The theoretical study can be classified into two scenarios,
the final state effect associated with the QGP fluid expansion and the initial state effect without the QGP formation.
In the final state scenario, the collective flow is related to the initial state geometry through the non-linear evolution, where hydrodynamic or kinetic calculations can roughly reproduce the data with tuned parameters~\cite{Shen:2016zpp,Bozek:2011if,Bzdak:2013zma,Qin:2013bha,Nagle:2013lja,Werner:2013tya,Werner:2013ipa,Bozek:2013ska,Schenke:2014zha,Bozek:2014cya,Bozek:2015swa,Zhou:2015iba,Weller:2017tsr,Mantysaari:2017cni,Zhao:2017rgg,Zhao:2020pty,Zhao:2020wcd,PHENIX:2018lia,Schenke:2020mbo,Schenke:2019pmk,OrjuelaKoop:2015etn,Bozek:2015qpa,PHENIX:2022nht}.
In contrast, the initial state effect describes the observed anisotropies by momentum correlation of the initially produced particles in the color field domain. One typical model is the Color Glass Condensate(CGC), which can qualitatively describe the experimental measurements such as two-particle and multi-particle correlations, long-range rapidity correlations and mass ordering~\cite{Schenke:2021mxx,Dusling:2012cg,Dusling:2012iga,Kovner:2012jm,Kovchegov:2012nd,Lappi:2015vta,Schenke:2015aqa,Dusling:2017dqg,Dusling:2017aot,Mace:2018yvl,Schenke:2016lrs}.
Compared to the final state scenario, a major difference is that the correlations are expected to be weaker for larger collision systems with more uncorrelated domains involved.

It was believed that comparative runs of p/d/$^3$He + Au collisions with variation of the initial state geometry
could provide useful information to identify the above two scenarios for flow-like signals in the small systems~\cite{PHENIX:2018lia}.
More specifically, with the initial state geometry dominated by the nucleon position fluctuations, models such as  MC-Glauber show an eccentricity ordering as: $\e_2^{\rm p+Au} < \e_2^{\rm d+Au} \simeq \e_2^{\rm {}^3He+Au}$, $\e_3^{\rm p+Au} \simeq \e_3^{\rm d+Au} < \e_3^{\rm {}^3He+Au}$. Hydrodynamic evolution  responses  directly to such initial state eccentricities and generates the associated flow ordering as: $v_2^{\rm p+Au} < v_2^{\rm d+Au} \approx v_2^{\rm {}^3He+Au}$, $v_3^{\rm p+Au} \approx v_3^{\rm d+Au} < v_3^{\rm {}^3He+Au}$~\cite{Nagle:2013lja,Welsh:2016siu,Shen:2016zpp}.
In contrast, CGC model calculations are not sensitive to such initial state geometry, predicting the flow anisotropies of much smaller magnitude for the three collision systems as:
$v_2^{\rm p+Au} \gtrsim v_2^{\rm d+Au} \gtrsim v_2^{\rm {}^3He+Au}$, $v_3^{\rm p+Au} \gtrsim v_3^{\rm d+Au} \gtrsim v_3^{\rm {}^3He+Au}$~\cite{Mace:2018yvl, Mace:2018vwq}.

\begin{table*}[thp]
\caption{\label{tab:parameter}
Parameters setups}
\begin{ruledtabular}
\begin{tabular}{cccccccccccc}
&\textrm{Parameter Set}&
\textrm{$p$}&
\textrm{$k$}&
\textrm{$n_c$}&
\textrm{$\omega$}&
\textrm{$\nu$}&
\textrm{$\tau_0$}&
\textrm{$(\eta/s)_{\text{min}}$}& \textrm{$(\eta/s)_{\text{slope}}$}&
\textrm{$(\zeta/s)_{\text{max}}$}&
\textrm{T$_{\text{switch}}$}\\
\colrule
I  & Sub-Nucl. Fluc.  & 0  & 0.6  & 5  & 0.5   & 0.2 & 0.6 & 0.28  & 1.6  & 0.022  & 154\\ 
II  & Nucl. Fluc.  & 1  & 1.6  & 1  & 0.4   & - & 0.6 & 0.09  & 1.0  & 0.0  & 151\\ 
III  & Sub-Nucl. Fluc.  & 0  & 0.28  & 6  & 0.92   & 0.55  & 0.37\footnote{with free-streaming}  & 0.11  & 1.6  & 0.032  & 151\\ 
\end{tabular}
\end{ruledtabular}
\end{table*}

With this motivation, the PHENIX Collaboration has measured the flow anisotropies of  $v_2(p_T)$ and $v_3(p_T)$ in p/d/$^3$He+Au collisions using the event-plane method \cite{PHENIX:2018lia} and the ``3 $\times$ 2PC" method \cite{PHENIX:2021ubk}, respectively. Both methods observed similar flow ordering: $v_2^{p+\text{Au}} < v_2^{d+\text{Au}} \approx v_2^{^3\text{He}+\text{Au}}$ and $v_3^{p+\text{Au}} \approx v_3^{d+\text{Au}} < v_3^{^3\text{He}+\text{Au}}$ \cite{PHENIX:2018lia}, which consists with the hydrodynamic predictions with MC-Glauber initial conditions~\cite{Nagle:2013lja,Welsh:2016siu,Shen:2016zpp,Zhao2023}. The PHENIX measurements also largely ruled out the CGC calculations with only initial state effects, which show different $v_2$ and $v_3$ orderings of smaller magnitude~\cite{PHENIX:2021ubk,Mace:2018vwq,Nagle:2018ybc}. Recently, the STAR collaboration  has also measured the flow coefficients of these three small systems using non-flow subtraction methods based on the template fit and the Fourier expansion fit, respectively~\cite{Lacey:2020ime,STAR:2022pfn}. 
Compared to the PHENIX results, the STAR measurement gave much larger $v_3(p_T)$ in p+Au and d+Au systems with the TPC detector at mid-rapidity. 
Such system independent $v_3$ data might indicate the existance of sub-nucleon fluctuation, while the discrepancy between measurements may also originate from different non-flow subtraction or different detector pseudorapidity acceptance~\cite{Nagle:2021rep,Lim:2019cys,PHENIX:2021ubk}.

On the theoretical side, the flow signals of these small collision systems have been studied by various models with different initial conditions \cite{Shen:2016zpp,Zhao:2022ugy,Schenke:2019pmk,Habich:2014jna,Bozek:2014cya,Braun:2019yjl}. The hydrodynamic calculations with MC-Glauber initial conditions can well describe the $v_2(p_T)$ and $v_3(p_T)$ data from PHENIX, but cannot explain the flow discrepancy between STAR and PHENIX, even with the longitudinal decorrelations in the 3+1-d simulations~\cite{Zhao:2022ugy}. Using the initial conditions with sub-nucleon fluctuations, hydrodynamic simulations produce similar $v_3$ values in p/d/$^3$He+Au collisions, which is obviously different from the PHENIX results and the calculations with nucleon fluctuations. However, these hydrodynamic calculations cannot simultaneously  describe the  STAR $v_2$ and $v_3$ data and the initial sub-nucleon structure is still not well constrained.

In this paper, we study the flow observables in p/d/$^3$He + Au collisions at $\sqrt{s_{NN}} = 200$~GeV, using 2+1-d \\
\vishnu~ model with \trento~ initial conditions including both nucleon and sub-nucleon flucuations. We tune the model parameters to fit the elliptic and triangular flow data from PHENIX and STAR, respectively, and calculate the 4-particle cumulants $c_2\{4\}$ in p+Au and d+Au systems.  We also evaluate the validity of the hydrodynamics by Knudsn numbers and draw the hydrodynamic predicted $v_3/v_2(p_T)$ band in p/d/$^3$He + Au collisions with the relatively reliable region of 2+1-d hydrodynamics.

\begin{figure}[tb]
  \center
    \includegraphics[
    width=0.9\columnwidth]{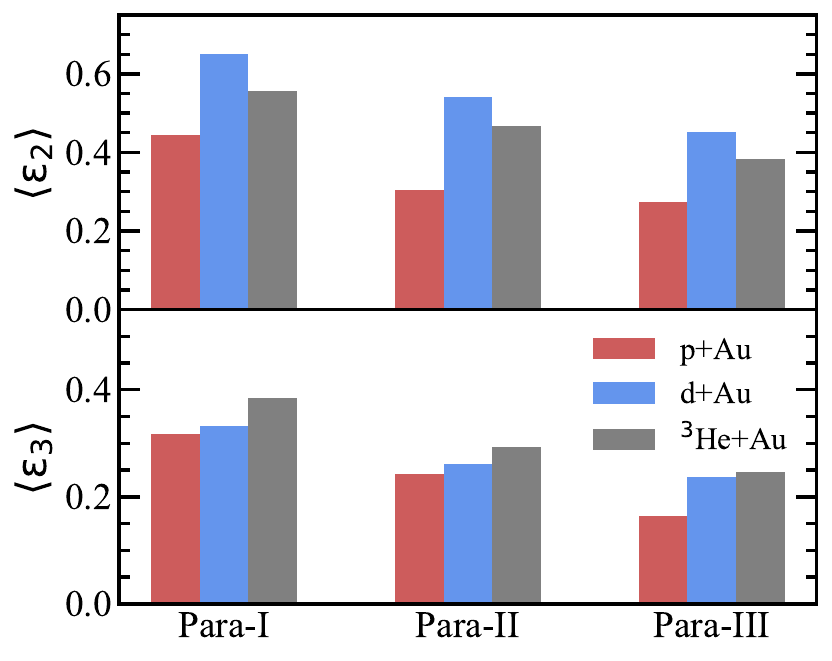}
    \caption{
    \label{fig:1}
    The initial eccentricities $\e_2$  and $\e_3$, calculated from {\trento} with impact parameter b $<$ 2 fm.
    }
\end{figure}
  
\begin{figure*}[t]
\center
  \includegraphics[
    width=0.9\textwidth]{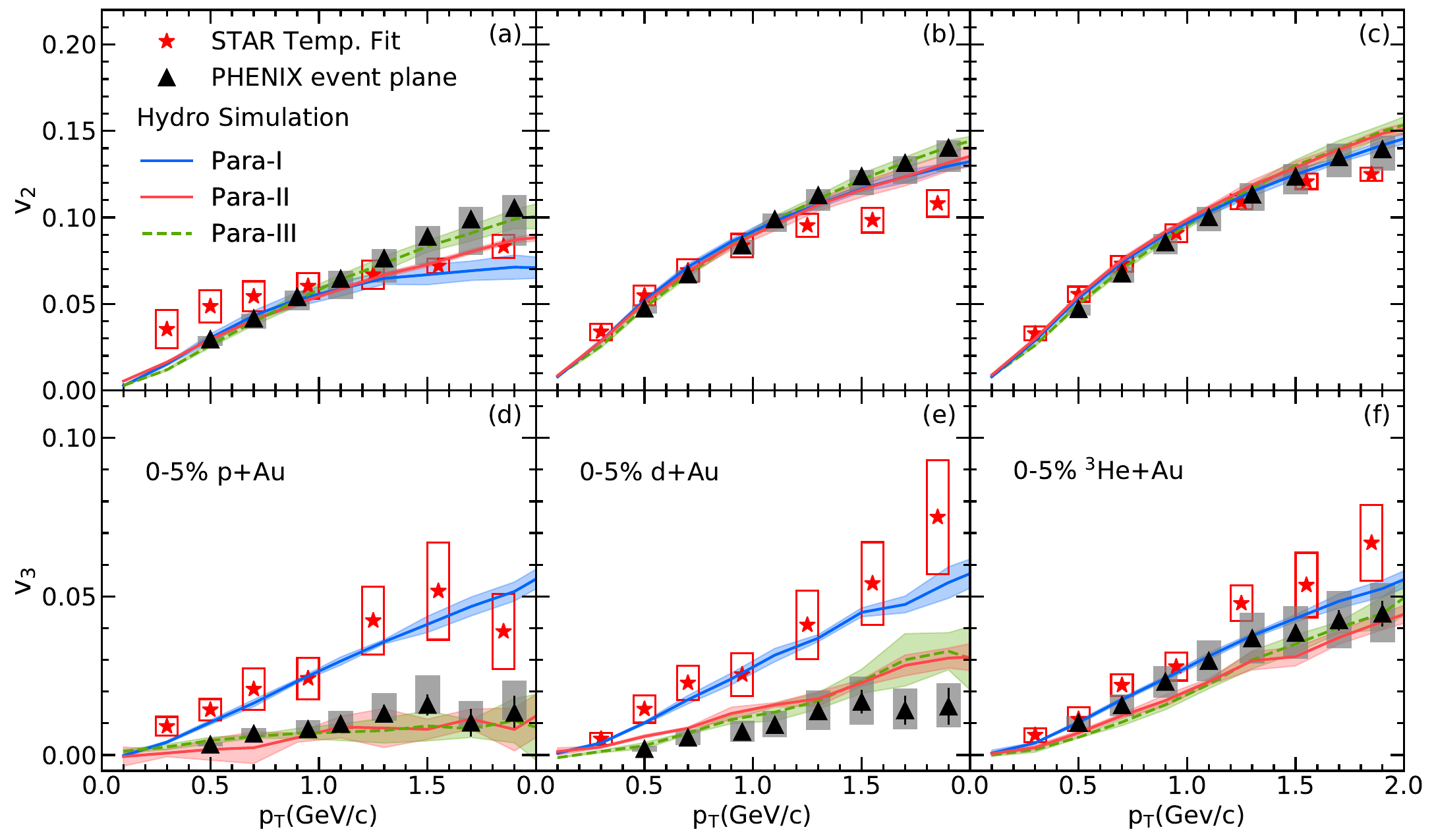}
    \caption{
    \label{PHENIXv23}
    Flow harmonics $v_2(p_T)$ and $v_3(p_T)$ for all charged hadrons in 0-5\% $p+$Au (a), $d+$Au (b) and $^3$He+Au (c) collisions at $\sqrt{s_{NN}} = 200$~GeV. The theoretical curves are
    calculated by iEBE-VISHNU using \trento~initial condition with three parameter sets listed in Table \ref{tab:parameter}. The STAR and PHENIX data are taken from \cite{Lacey:2020ime} and \cite{PHENIX:2018lia},respectively.
    }
\end{figure*}

\section{\label{sec:model}MODEL SETUP}
In this paper, we implement iEBE-VISHNU to study the flow observables in  p/d/$^3$He+Au collisions at $\sqrt{s_{NN}}=200$ GeV. iEBE-VISHNU~\cite{Shen:2014vra} is an event-by-event hybrid model that combines 2+1-d viscous hydrodynamics VISH2+1~\cite{Heinz:2005bw,Song:2007fn,Song:2007ux} for the QGP evolution, a particle sampler iSS~\cite{Song:2010aq,Song:2010mg} for the particlization at a switching temperature,
and the hadron cascade model UrQMD~\cite{Bass:1998ca,Bleicher:1999xi} for the subsequent hadronic evolution. Following~\cite{Xu:2016hmp,Zhao:2017yhj,Zhao:2018lyf,Zhao:2017rgg,Zhao:2020pty}, we use the HotQCD+HRG equation of state (EoS)~\cite{Huovinen:2009yb,HotQCD:2014kol,Bernhard:2016tnd} as input, and set temperature-dependent specific shear viscosity $\eta/s$ and bulk viscosity $\zeta/s$ \cite{Bernhard:2019bmu}.

We implement {\trento}, a parameterized initial condition model, to generate the initial entropy density for the hydrodynamic simulations starting at $\tau_0$ ~\cite{Moreland:2014oya,Bernhard:2016tnd,Moreland:2018gsh}.
In the case without sub-nucleon structure, the fluctuations come from the distribution of nucleon center position. For each nucleon, its density distribution is parameterized as a Gaussian function with nucleon Gaussian width $\omega$:
\begin{equation}\label{eq:density1}
  \rho_{\text{nucleon}}(\textbf{x}) = \frac{1}{(2\pi \omega^2)^{3/2}} \text{exp}(-\frac{\textbf{x}^2}{2\omega^2}),
\end{equation}

On the other hand, when considering the sub-nucleon fluctuation, the nucleon is assumed to be composed of independent constituents, and the nucleon density is written as:
\begin{equation}
  \rho_{\text{nucleon}}(\textbf{x}) = \frac{1}{n_c}\sum_{i=1}^{n_c} \rho_{\text{constit}} (\textbf{x} - \textbf{x}_i),
\end{equation}
where $n_c$ is the constituent number, $\textbf{x}_i$ is the position of the i-th constituent, density $\rho_{\text{constit}}$ defined as $\rho_{\text{constit}} (\textbf{x}) = \frac{1}{(2\pi v^2)^{3/2}} \text{exp}(-\frac{\textbf{x}^2}{2v^2})$. The constituent Gaussian width $v$ relates to the nucleon width $\omega$ with a standard deviation r: $\omega = \sqrt{r^2 + v^2}$ and in this case $\omega$ is defined as the root mean square radius of a nucleon.

After obtaining the nucleon density distribution, the fluctuated thickness of the colliding nucleons is then written as:
\begin{equation}\label{eq:thick2}
  \tilde{T}_{A,B}(\textbf{x})\equiv \int dz\ \frac{1}{n_c}\sum^{n_c}_{i=1}\gamma_i \rho_{\text{constit}}(\textbf{x} - \textbf{x}_i \pm \textbf{b}/2).
\end{equation}
Here, besides the nucleon/constituent position, the initial fluctuation is controlled mainly by the gamma random variable $\gamma_i$, which is parameterized by the shape factor k. The resulting standard devaition of the initial  fluctuation is denoted as
\begin{equation}
  \sigma_{\rm fluct} = 1/ \sqrt{k n_c}.
\end{equation}
With the fluctuating thickness $\tilde{T}_{A,B}$,  the initial entropy density at mid-rapidity can be calculated by the generalized means with a dimensionless parameter $p$:
\begin{equation}
   \frac{dS}{d^2x_{\perp}d\eta}\bigg{|}_{\eta=0} \propto \bigg{(} \frac{\tilde{T}_A + \tilde{T}_B}{2} \bigg{)}^{1/p}
\end{equation}

\footnotetext{with free-streaming}
Tab.~\ref{tab:parameter} list the model parameters used in the calculations for p+Au, d+Au and $^3$He+Au collisions at $\sqrt{s_{NN}}$ = 200 GeV. The Para-I is tuned to fit the $v_2(p_T)$ and $v_3(p_T)$ data from STAR, which includes the initial nucleon sub-structure and a small constituent width to enlarge the fluctuation.
The Para-II and Para-III fit the published PHENIX data, which are tuned with/without sub-nucleon fluctuation, respectively. Para-III with sub-nucleon structure is similar to the Bayesian analyses in the p+Pb and Pb+Pb collisions~\cite{Moreland:2018gsh}, except that  \trento~paratemr k and $\nu$ are tuned to fit the charged particle multiplicity distribution in d+Au collisions at $\sqrt{s_{NN}} = 200$~GeV.
Note that, reproducing the STAR $v_2$ and $v_3$ at the same time requires a large shear viscosity in Para-I, which lies outside the usual parameter range of the hydrodynamic approaches. This will be further discussed in
section IV. Fig.~\ref{fig:1} plots the averaged eccentricities $\varepsilon_{2,3}$,  calculated from the {\trento} initial conditions with the parameter sets in Tab.~\ref{tab:parameter}.  Compared to the early results of the MC-Glauber model~\cite{Nagle:2013lja,PHENIX:2018lia}, Fig.~\ref{fig:1} shows a weaker  ordering of $\e_n$  for p+Au, d+Au and $^3$He+Au collisions. Such weaker ordering in Para-I and Para-III can be explained by sub-nucleonic fluctuations. Due to the imprinted multiplicity fluctuations, Para-II without subnucleonic fluctuations also shows a weaker ordering of $\e_n$ for the {\trento} initial conditions~\cite{Welsh:2016siu}.

\begin{figure*}[t]
  \center
    \includegraphics[
    width=0.9\textwidth]{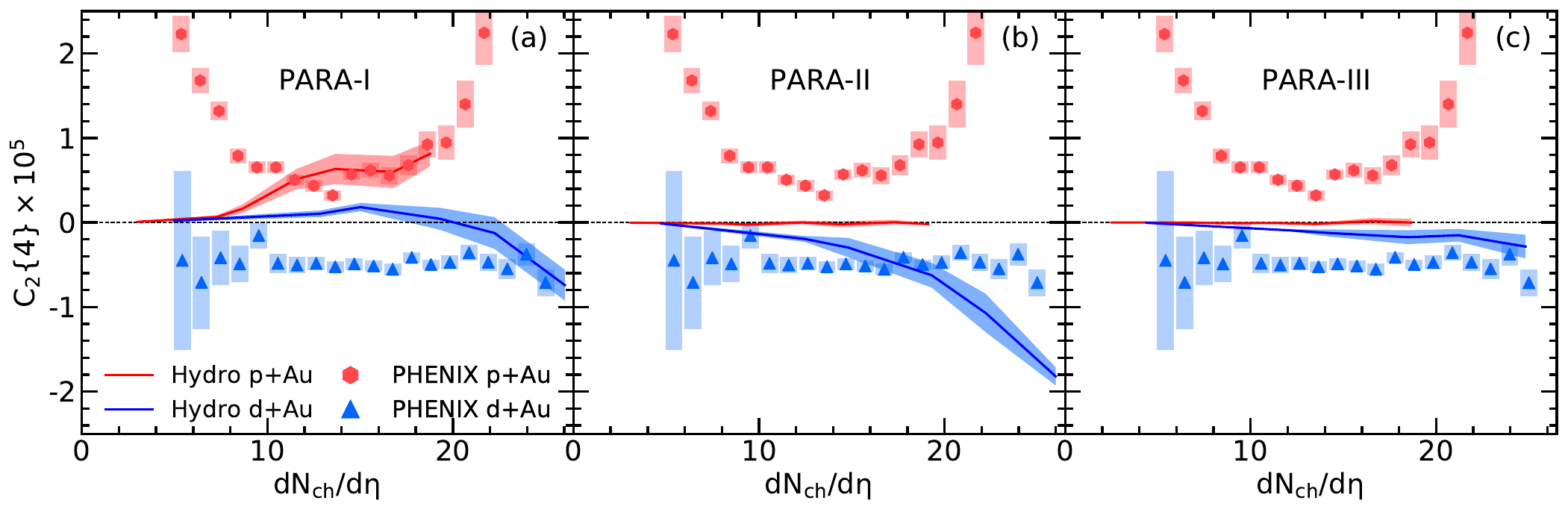}
    \caption{
    \label{c24}
    $c_2\{4\}$ as a function of $dN_{\text{ch}}/d\eta$ in $p$+Au and $d$+Au collisions at $\sqrt{s_{NN}} = 200$~GeV, calculated by iEBE-VISHNU with \trento~initial condition with three parameter sets. The PHENIX data are taken from ~\cite{PHENIX:2017xrm}.
    }
\end{figure*}

\section{\label{sec:results}Results and discussions}

In this section, we show the flow harmonic results calculated by iEBE-VISHNU with the parameters listed in Table ~\ref{tab:parameter}. All parameter sets in Table~\ref{tab:parameter} are well tuned to reproduce the multiplicity distribution in d+Au collisions.  We also tune Para-I to fit $v_2(p_T)$ and $v_3(p_T)$ from STAR, and Para-II and Para-III to fit $v_2(p_T)$ and $v_3(p_T)$ from PHENIX.  

Figure~\ref{PHENIXv23} shows the differential flow harmonics $v_2(p_T)$ and $v_3(p_T)$ of all charged hadrons in 0-5\% $p$+Au, $d$+Au and ${}^3$He+Au collisions~\footnote{In our hydrodynamic simulation, we follow the PHENIX centrality definition, and calculate the flow harmonics in p/d/$^3$He+Au collisions with 0-5\% centrality cut. Note that the STAR measurements focus on the Ultra-Central(UC) p+Au collisions with 0-2\% centrality and the most central d/$^3$He+Au  collisions with 0-10\% centrality. As argued  by the STAR paper \cite{STAR:2022pfn}, the orderings of flow harmonics are insensitive to the centrality definition. This is also confirmed by our hydrodynamic simulations. }.  For the triangular flow $v_3(p_T)$,  the STAR and PHENIX measurements are not consistent, particularly for p+Au and d+Au collisions, where the STAR data are larger than the PHENIX data by a factor of 3. Such apparent discrepancies may be due to the different rapidity region and non-flow substraction methods used by these two collaborations~\cite{Lacey:2020ime,PHENIX:2021ubk}. Therefore, we fit the STAR and PHENIX data, respectively.  Following \cite{Bilandzic:2010jr,Zhou:2015iba}, we use the two-subevent cumulant method to calculate the two particle correlations with a kinematic cut $0.2<p_T<2.0$ GeV/c and $|\eta|<1.0$ with a gap $|\Delta \eta|>1.0$.

With the Para-I, iEBE-VISHNU nicely fits the $v_2(p_T)$ and $v_3(p_T)$ data measured by STAR. We find that sub-nucleon fluctuations are essential to produce larger $v_3$, which are insensitive to the collision systems variation.  Meanwhile a large shear viscosity is also required to simultaneously fit the $v_2$ and $v_3$ data from STAR. The validity of hydrodynamic simulations with such large shear viscosities will be discussed in  the next section. Note that the effects of sub-nucleon fluctuations on the flow of small systems also have been studied and discussed in the earlier paper~\cite{Schenke:2019pmk}. For the PHENIX measurements, iEBE-VISHNU simulations with nucleon fluctuations in the initial state (Para-II) were able to roughly reproduce the $v_2$ and $v_3$ data within the statistical error bars. Meanwhile, one can achieve similar results with sub-nucleonic initial conditions with a free streaming (Para-III). Here, Para-III is the one obtained from the Bayesian analysis for Pb+Pb and p+Pb collisions at $\sqrt{s_{NN}} = 5.02$~TeV \cite{Moreland:2018gsh} except for the fluctuation parameter k and the constituent width $\nu$, which are tuned to fit the multiplicity fluctuation in top RHIC energy.
We conclude that, for the PHENIX measurements, iEBE-VISHNU simulations with  both nucleonic and sub-nucleonic initial state fluctuations can fit the $v_2$ and $v_3$ hierarchies in p/d/$^3$He+Au collisions.

Fig.~\ref{c24} plots the 4-particle cumulant $c_2\{4\}$ as a function of $dN_{\rm ch}/d\eta$ for $p$+Au and $d$+Au collisions at $\sqrt{s_{NN}} = 200$~GeV. Panel (a) shows the iEBE-VISHNU predictions with the Para-I, which is tuned to fit the STAR $v_2(p_T)$ and $v_3(p_T)$ data and generates positive $c_2\{4\}$ for p+Au collisions and negative $c_2\{4\}$ for d+Au collisions in the high multiplicity events, consistent with the experimental measurements qualitatively.
In fact, large event-by-event fluctuation in Para-I leads to positive $c_2\{4\}$ in p+Au collisions.
While for the d+Au collisions, the intrinsic geometry of the deutron gives a dominant contribution to the initial eccentricities, leading to a negative $c_2\{4\}$ in the high multiplicity events.
Panels (b) and (c) show the iEBE-VISHNU results, calculated with the Para-II and Para-III fitting the PHENIX $v_2(p_T)$ and $v_3(p_T)$ data.  For p+Au collisions, $c_2\{4\}$ is always close to zero for both parameter sets over the whole range of multiplicities, due to small flow fluctuations. For d+Au collisions, $c_2\{4\}$  are always negative due to the intrinsic geometry of the deutron.

\begin{figure*}[t]
  \includegraphics[width=1.0\textwidth]{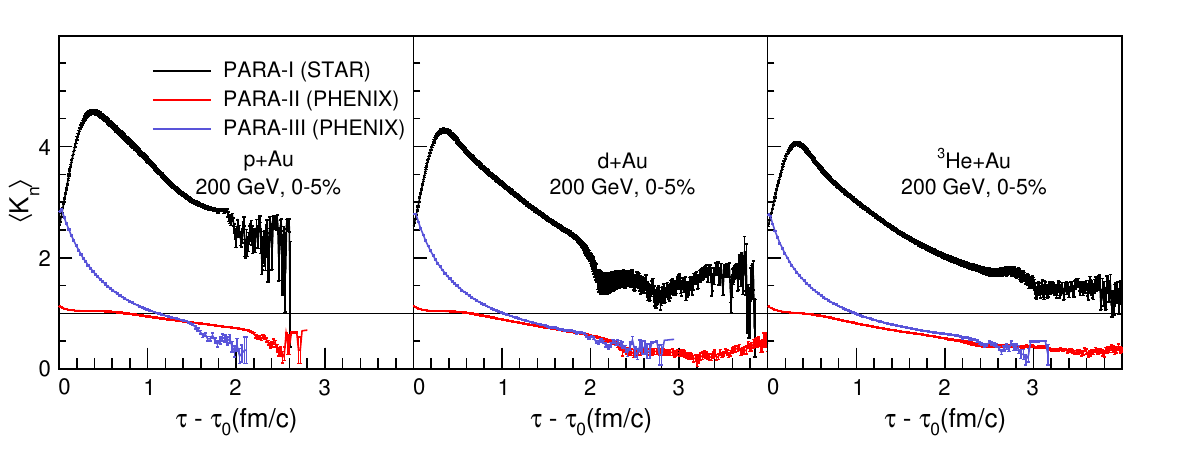}
  \caption{
  \label{RKn} The evolution of the average Knudsen number $\langle K_n \rangle$ within the freezeout hypersurface for  0-5\% p/d/$^3$He+Au collisions.
  }
\end{figure*}

\section{Applicability of hydrodynamic simulation }

We have noticed that in the above calculation, the specific shear viscosity in
some parameter sets tuned to fit the $v_2(p_T)$ and $v_3(p_T)$  data become quite large. In order to evaluate the validity of the hydrodynamic simulations in small systems, we calculate the Knudsen number $K_{n}$
defined as~\cite{Niemi:2014wta}:
\begin{align}\label{eq:kn}
  K_n = \tau_{\pi}\theta=5\frac{\eta\theta}{sT},
\end{align}
where $\tau_{\pi}$ is the relaxation time associated with the microscopic time scale and $\theta = \partial_{\mu} u^\mu$ is the expansion rate associated with the macroscopic hydrodynamic time scale. $K_n \rightarrow 0$ is the perfect fluid limit where the local equilibrium is maintained during the hydrodynamic evolution.  $K_n \rightarrow \infty$ is the other limit, which corresponds to the case that the fluid system breaks up into free-streaming particles. It is generally suggested that the hydrodynamics is relatively reliable with $K_n < 1$~\cite{Niemi:2014wta,Bouras2009}
\footnote{Besides the expansion rate $\theta$ defined in Eq.~\ref{eq:kn}, the macroscopic scale can also be estimated from other macroscopic gradients~\cite{Niemi:2014wta}. For the propose of illustration, we use Eq.~\ref{eq:kn} and set the criterion $\langle K_n \rangle > 1$ for the failure of hydrodynamics, which is associated with the fact that the macroscopic expansion rate is larger than the microscopic relaxation rate.}.

Fig.~\ref{RKn} shows the time evolution of the averaged Knudsen number $\langle K_n \rangle$ in the event-by-event hydrodynamic simulations for p/d/$^3$He+Au collisions at 0-5\% centrality. The average is taken within the freeze-out hypersurface with the local energy density as the weight for each time step.  
For the Para-I with sub-nucleon fluctuations 
we observed that the  averaged Knudsen number $\langle K_n \rangle$ is always larger than 1 throughout the whole evolution for different collision systems.
Obviously, such a large Knudsen number indicates that the hydrodynamic simulations are beyond their applicable limit, which is mainly due to the large specific shear viscosity $\eta/s \sim 0.28$ and the large initial gradients introduced by fluctuations to fit the $v_3$ data.
In contrast, the average Knudsen number for Para-II is about or less than 1 with a smaller specific shear viscosity $\eta/s \sim 0.09$. For Para-III, the Knudsen number lies between those of  Para-I and Para-II, which is large in the early time due to the free streaming evolution before thermalization, but drops below 1 after certain time of hydrodynamic evolution.
In short, Fig.~\ref{RKn} suggests that hydrodynamic simulations with Para-I that tuned to fit the STAR data are beyond the limit due to the large Knudsen number.

\begin{figure*}[t]
  \center
    \includegraphics[
    width=0.9\textwidth]{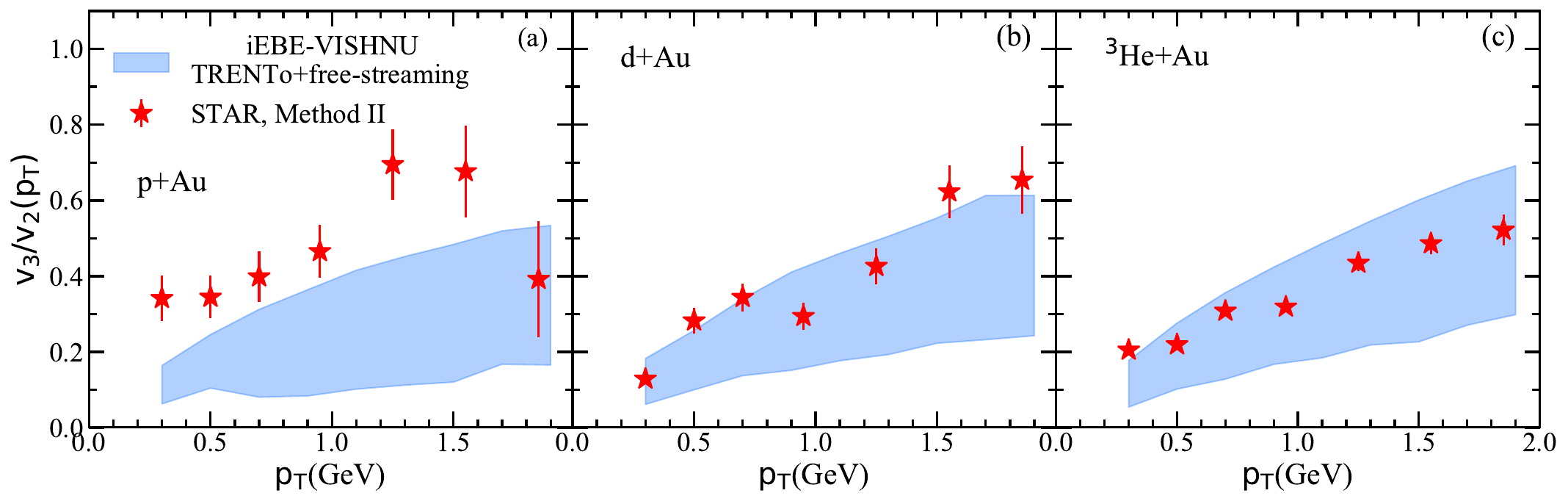}
    \caption{
    \label{band} The ratio of $v_3(p_T)$ and $v_2(p_T)$ for all charged hadrons in 0-5\% p/d/$^3$He+Au collisions, calculated by iEBE-VISHNU using the the parameter range listed in Tab.~\ref{tab:parameterII}, together with the constraint  $\langle K_n \rangle<1$. The STAR data are taken from \cite{Lacey:2020ime}.
    }
\end{figure*}

\begin{table}[h]
\caption{\label{tab:parameterII}%
Free parameter range for the hybrid model.
}
\begin{ruledtabular}
\begin{tabular}{llllll}
\textrm{Parameter}&
\textrm{Description}&
\textrm{Prior range} &\\
\colrule
$\omega$ & Nucleon width & [0.4, 1.0] fm\\
$\nu$ & Constituent width ($< \omega$) & [0.3, 1.0] fm \\
$n_c$ & Number of nucleon constituents & [1, 9] \\
$\tau_{\text{fs}}$ & Free-streaming time & [0.2, 0.8] fm/c \\
$T_{\text{switching}}$ & Switching temperature & [135, 165] MeV \\
Free-stream && on/off \\
Sub. fluct. && on/off \\
\end{tabular}
\end{ruledtabular}
\end{table}

To further investigate whether iEBE-VISHNU could fit all the experimental data within its hydrodynamic limit, we explore the model parameter space as far as possible but with the constraint of the Knudsen number $\langle K_n \rangle<1$ at the end of the evolution. Our test parameter sets correspond such initial conditions with/without nucleon substructure and with/without the free-streaming effect. The range of the free parameters is listed in Tab.~\ref{tab:parameterII}. With $n_c=1$, the initial conditions include only nucleon fluctuations and  with $n_c=2-9$, the initial conditions include sub-nucleon fluctuations. In our investigation, the effective shear viscosity $\eta/s$ and the shape parameter k are fixed to reproduce the $v_2(p_T)$ data in 0-5\% $^3$He+Au collisions and the multiplicity distribution of d+Au collisions with neglecting the bulk viscosity. 

Fig.~\ref{band} shows the $p_T$ dependent $v_3(p_T)/v_2(p_T)$ ratio in 0-5\% p/d/$^3$He+Au collisions, where the theoretical band is calculated by iEBE-VISHNU with the parameter range listed in Tab.~\ref{tab:parameterII}, together with the constraint  $\langle K_n \rangle<1$.
The experimental data are taken from STAR with the statistical uncertainty of $v_2(p_T)$ and $v_3(p_T)$ using the error propagation formula.
As shown in  panels (b) and (c), the flow harmonic ratio $v_3(p_T)/v_2(p_T)$ in d/$^3$He+Au collisions can be reproduced by iEBE-VISHNU within the allowed parameter range $\langle K_n \rangle<1$. While panel (a) shows that the upper limit of the $v_3(p_T)/v_2(p_T)$ ratio in p+Au collisions, calculated from iEBE-VISHNU simulations,
is clearly below the experimental data. These results  indicate that the current hybrid model calculations are not able to simultaneously describe the STAR flow data in the three small collision systems with reasonable parameter range within the hydrodynamic limit.

\section{\label{sec:summary}SUMMARY}
In this paper, we implemented  iEBE-VISHNU  with \trento ~initial condition to study the collective flow in p/d/$^3$He+Au collisions at $\sqrt{s_{NN}} = 200$~GeV.
For the PHENIX measurements,  $v_2(p_T)$ and $v_3(p_T)$ data show obvious hierarchies for different collision systems, which can be reproduced by our hybrid model simulations with nucleon/sub-nucleon fluctuating initial conditions.
The related simulations also reproduce a negative 4-particle cumulant $c_2\{4\}$ for d+Au collisions, but give an almost zero $c_2\{4\}$ for p+Au collisions, which can not describe the positive $c_2\{4\}$ measured by PHENIX.
For the STAR measurements, the magnitude of $v_3$ are insensitive to the collision systems.
iEBE-VISHNU simulations with sub-nucleon fluctuating initial conditions can fit these $v_2$ and $v_3$ data, which can also roughly reproduce the positive and negative $c_2\{4\}$  measured in the high multiplicity p+Au and d+Au collisions, respectively. 
However, due to the large shear viscosity applied to fit experimental data, the hydrodynamic simulations has already beyond its limits with the average Knudsen number $\langle K_n \rangle$ obviously larger than one for these three collision systems. We also explore the model parameter space as far as possible with the constraint of the Knudsen number $\langle K_n \rangle<1$, and found that \vishnu~ with the \trento~ initial condition with/without sub-nucleon fluctuations cannot describe the  experimental measured $v_3 / v_2$ ratio for p+Au collisions.

Our calculations demonstrate that for a meaningful evaluation of the collective flow in the small systems, one should also evaluate the validity of hydrodynamics.
As the collision system become smaller, the isotropization and thermalization conditions are harder and harder to reach.
Besides applying a full 3+1-d hydrodynamic simulations, other improved hydrodynamic theories like anisotropic hydro~\cite{ Bazow:2013ifa,McNelis:2021zji, Molnar:2016vvu, Florkowski:2010cf} should be implemented to the small systems that may not reach equilibrium in the early stage.
It was also found that the fragmentation and mini-jets effects become more important in small collision systems~\cite{Zhao:2020wcd,Note1}, a comprehensive model includes the core-corona effects~\cite{Werner:2007bf,Kanakubo:2019ogh} is also required to further evaluate the flow signals in p/d/$^3$He+Au collisions at $\sqrt{s_{NN}} = 200$~GeV.


\begin{acknowledgments}
  We thank Wenbin Zhao for helpful discussions.
  This work was supported in part by the NSFC under grant No.~12247107, No.~12075007 and No.~12147173 (B.F.)
  We also acknowledge the extensive computing resources provided by the Supercomputing Center of Chinese Academy of Science (SCCAS), Tianhe-1A from the National Supercomputing Center in Tianjin, China and the High-performance Computing Platform of Peking University.
\end{acknowledgments}

\appendix

\bibliography{apssamp}

\end{document}